\documentclass[11pt]{article}

\usepackage{graphicx}
\usepackage{epstopdf}
\usepackage{amsmath}
\usepackage{amssymb}
\usepackage{amsfonts}
\usepackage{amsthm}
\usepackage[usenames]{color}
\usepackage{array}

\usepackage[
colorlinks=true,
linkcolor=blue,
urlcolor=blue,
filecolor=blue,
citecolor=red,
pdfstartview=FitV,
pdftitle={},
pdfauthor={},
pdfsubject={},
pdfkeywords={},
pdfpagemode=None,
bookmarksopen=true
]{hyperref}

\usepackage{epsfig}

\usepackage{hyperref}

\newcommand{\bea}{\begin{eqnarray}}
\newcommand{\eea}{\end{eqnarray}}
\newcommand{\ba}{\begin{array}}
	\newcommand{\ea}{\end{array}}
\newcommand{\ee}{\end{equation}}
\numberwithin{equation}{section}

\begin{document}

\begin{flushright}
	\texttt{\today}
\end{flushright}

\begin{centering}
	
	\vspace{2cm}
	
	\textbf{\Large{
			  Dominant Spacetime  in Three Dimensional de Sitter Gravity  }}
	
	\vspace{0.8cm}
	
	{\large   Reza Fareghbal$^{1}$, Vahid Reza Shajiee$^{2}$ }
	
	\vspace{0.5cm}
	
	\begin{minipage}{.9\textwidth}\small
		\begin{center}
			
			{\it  $^{1}$Department of Physics,
				Shahid Beheshti University, 1983969411,
				 Tehran , Iran \\
				  $^{2}$Young Researchers and Elite Club, Mashhad Branch, Islamic Azad University, Mashhad, Iran } \\			
			
			\vspace{0.5cm}
			{\tt   r$\_$fareghbal@sbu.ac.ir, v.shajiee@mshdiau.ac.ir}
			\\
			
					\end{center}
	\end{minipage}

	%\end{center}

	\begin{abstract}
	In three dimensions, Kerr-de Sitter spacetime as a solution of Einstein gravity with positive cosmological constant has a single cosmological horizon. The flat-space limit (zero cosmological constant limit) of this spacetime is well-defined and yields  the flat-space cosmological solution which is a significant  spacetime in the context of flat-space holography. In this paper, we calculate the free energy of this spacetime and compare it with the free energy of the three-dimensional de Sitter spacetime. We investigate which one of these two spacetimes will dominate in the semi-classical approximation for estimating the partition function . It is shown that for the same temperature of cosmological horizon of two spacetimes this is the  de Sitter spacetime which is always dominant. Hence, contrary to  asymptotically flat and asymptotically AdS spacetimes, there is no  phase transition in three dimensional de Sitter gravity.

	\end{abstract}

\end{centering}

\newpage

%\tableofcontents

%\setcounter{equation}{0}

\section{Introduction}
Einstein gravity in three dimensions with a positive cosmological constant does not have black hole solutions. Nevertheless, solutions of the equations of motion can be obtained that have only a single cosmological horizon. These solutions are known as three dimensional Kerr-de Sitter (KdS)  spacetimes \cite{Park:1998qk}. For the cosmological horizon of these spacetimes, temperature and entropy can be defined similar to black holes. It is also possible to find a coordinate transformation that locally transforms these solutions into de Sitter (dS) spacetime. Since the periodicity of the periodic coordinates in the KdS metric, which is obtained from the transformation of the coordinates on the dS spacetime, is different from $2\pi$, these spacetimes are also called three-dimensional conical defects \cite{Balasubramanian:2001nb}. Moreover, taking the flat-space limit (zero cosmological constant limit ) from these solutions leads to flat-space cosmological (FSC) solutions \cite{Cornalba:2002fi}, which are important spacetimes in flat-space holography \cite{Bagchi:2012xr}.

If we want to write the three-dimensional de Sitter gravity partition  function for an ensemble whose temperature is the same as the temperature of the cosmological horizon of ds$_3$,  then the KdS spacetimes with the same horizon temperature can also become important. In fact, in the semi-classical approximation we estimate the partition function with its saddle points. They are obtained from the on-shell action evaluated for the classical solutions of the equations of motion. In addition to the dS spacetime, it is possible that the KdS solutions  also play the role of the dominant term. Therefore, determining  the  dominant spacetime  in this approximation can be an important  problem in quantum gravity.

In this paper, we study this problem and calculate the free energy for both dS and KdS spacetimes using on-shell action. The condition that the horizon temperature of KdS and dS are equal results in a simplification in the expression related to the free energy of KdS so that these two free energies can be easily compared. The final result obtained from this comparison is that dS spacetime is always  preferable. The free energy we obtain for KdS correctly gives the thermodynamic quantities related to this spacetime, which is also a good indication of the correctness of the calculations.

It should be noted that when we talk about the dominant spacetime in de Sitter gravity, the Hawking-Page phase transition  and the critical temperature for this phase transition lose their meaning. The reason is that in order to calculate the free energy, we assume that the temperature of the horizons is equal to the de Sitter temperature, which is given by the de Sitter radius, $\ell$. This radius is determined by the cosmological constant which is fixed for the theory.

Investigating the aforementioned problem in higher dimensions presents challenges. This difficulty arises because asymptotically dS  black holes in higher dimensions possess both an event horizon and a cosmological horizon, each with distinct temperatures. This disparity raises questions about the existence of thermodynamic equilibrium for these solutions. Moreover, if we include de Sitter spacetime in an ensemble with a temperature matching that of the dS cosmological horizon,   whether they can be considered  part of the same ensemble as de Sitter spacetime remains a question. Various solutions have been proposed in recent years to address this issue
(see for example \cite{Carlip:2003ne}-\cite{Tannukij:2020njz}).

Our primary motivation for investigating this problem lies in the fact that taking the flat-space limit (zero cosmological constant limit) of dS and KdS spacetimes,  results in the three-dimensional flat and flat-space cosmological solution (FSC), respectively. Reference \cite{Bagchi:2013lma} has demonstrated a Hawking-Page phase transition between these two asymptotically flat spacetimes. Notably, the calculation in \cite{Bagchi:2013lma} involved a boundary term that is half of the usual Gibbons-Hawking-York term \cite{Gibbons:1976ue,York:1972sj}. In our current study, we employ the same boundary term and observe  that its inclusion does not lead to  KdS spacetimes  becoming  dominant.

The structure of this paper is as follows: In section 2, we introduce the KdS solutions and calculate their free energy. In  section 3, we compare the free energies of dS and KdS and determine the dominant spacetime. Last section is devoted to conclusion.

\section{Kerr-de Sitter spacetime and its free energy}

\subsection{Kerr-de Sitter spacetime}
In three dimensions, Einstein gravity with a positive cosmological constant has other solutions in addition to de Sitter solution, which is called Kerr-de Sitter (KdS) spacetime. This spacetime, which is given by 
\begin{equation}\label{3DKdSm}
ds^2=-\frac{(r^2+r_+^2)(r_-^2-r^2)}{\ell^2 r^2}dt^2+\frac{\ell^2 r^2}{(r^2+r_+^2)(r_-^2-r^2)}dr^2+r^2\left(d\phi-\frac{r_+r_-}{\ell r^2}dt\right)^2,
\end{equation}
 has a cosmological horizon which is located at $r=r_-$. Its mass, $M$, and angular momentum $J$ are given in the definition of $r_\pm$ as \footnote{In contrast  to \cite{Park:1998qk}, where the  radius of cosmological horizon is  $r=r_+$, we interchanged the definitions of $r_+$ and $r_-$ in this paper. Moreover, in order to have a well-defined flat-space limit we changed $M$ in \cite{Balasubramanian:2001nb} to $-M$.  }
 \begin{eqnarray}\label{r+r-}
  r_-^2 &=& 4 G M \ell^2\left(\sqrt{1+{J^2\over M^2\ell^2}}-1\right), \\
  \nonumber r_+^2 &=& 4 G M \ell^2\left(\sqrt{1+{J^2\over M^2\ell^2}}+1\right),
\end{eqnarray}
where $\ell$ is the radius of dS spacetime and is related to the cosmological constant as $\Lambda=2/\ell^{2}$. Unlike dS spacetime, KdS has a singularity at $r=0$. It is worth mentioning that one recovers  dS space time by setting $M=-1/8G$ and $J=0$ in metric \eqref{3DKdSm}.

Looking at metric \eqref{3DKdSm}, we see that KdS is very similar to BTZ black hole which is a solution of three dimensional Einstein gravity with negative cosmological constant. Similar to BTZ which is given by a local transformation from AdS$_3$, one can find a local change of coordinates which transforms dS$_3$ to KdS. If the periodicity of periodic coordinate in dS$_3$ is  $2\pi$, this coordinate transformation results in $2\pi\alpha$ for the period of $\phi$ in \eqref{3DKdSm}. Hence KdS is known as conical defect in literature \cite{Balasubramanian:2001nb}. 

It is possible to define temperature and entropy for the cosmological horizon $r_-$. They are given by \footnote{In the present work, the hatted quantities  $\hat T$ and  $\hat{\Omega}$   stand for the  Lorentzian solution  and  $T$ and $\Omega$ are used for the Euclidean signatures.}
\begin{equation}\label{tem and entropy}
  \hat T={\left(r_{-}^{2}+r_{+}^{2}\right)\over{2\pi r_{-}\ell^{2}}},\qquad S=\dfrac{2\pi r_-}{4G}.
  \end{equation}  
Moreover, the angular velocity of the horizon is given by $\hat{\Omega}=r_{+}/\ell r_{-}$.

One of the properties  of this metric, which is  the main reason for our attention to this spacetime, is that taking the flat-space  limit  leads to the flat-space cosmological solution (FSC), which is an important asymptotically flat spacetime  in the context of flat-space holography \cite{Bagchi:2012xr}. Not only metric, but also the flat-space limit of temperature and entropy in \eqref{tem and entropy} results in the corresponding quantities of FSC. FSC has a cosmological horizon which is located at the radius given by $\ell\rightarrow\infty$ limit of $ r_-$. It is worth mentioning that FSC can also be obtained  by taking the flat-space limit of the BTZ metric. However, it is essential to note that the  BTZ spacetime represent a black hole, characterized by non-cosmological horizons. Hence, a metric that lacks a cosmological horizon (BTZ metric), undergoes a transformation into a metric with a cosmological horizon (FSC metric) after taking the flat-space limit.  Indeed,   one of the intriguing features of the KdS metric   is that both before and after the flat-space limit, the horizons remain cosmological in nature. Consequently, FSC exhibits a closer resemblance to KdS rather than BTZ.

In the rest of this paper we need to work with Euclidean KdS. Its metric is given by performing the Wick rotation $t=i\tau$ together with $r_{+}=-ir_{p}$,
\begin{equation}\label{E3DKdSm}
ds_{E}^2=\frac{(r^2-r_p^2)(r_-^2-r^2)}{\ell^2 r^2}d\tau^2+\frac{\ell^2 r^2}{(r^2-r_p^2)(r_-^2-r^2)}dr^2+r^2\left(d\phi-\frac{r_p r_-}{\ell r^2}d\tau\right)^2.
\end{equation}
The definition of $r_p$ is equivalent to define  Euclidean mass $M_E$ and angular momentum $J_E$ as
\begin{equation}\label{Euc M J}
 M_E=-M,\qquad J_E=iJ.
 \end{equation} 
Thus, we can write $r_-$ and $r_p$ in terms of Euclidean parameters as\footnote{The flat space-limit of Euclidean KdS \eqref{E3DKdSm} with \eqref{Euclidean radius} is well-defined and results in the metric of Euclidean FSC introduced in \cite{Bagchi:2013lma}.}
\begin{eqnarray}\label{Euclidean radius}
  r_-^2 &=& 4 G M_E \ell^2\left(1-\sqrt{1-{J_E^2\over M_E^2\ell^2}}\right), \\
  \nonumber r_p^2 &=& 4 G M_E \ell^2\left(1+\sqrt{1-{J_E^2\over M_E^2\ell^2}}\right).
\end{eqnarray}

In order to check that there is no conical singularity on the horizon, one needs to take the near horizon limit of \eqref{E3DKdSm}. Plugging $r^{2}=r_{-}^{2}+\epsilon \rho^{2}$ into \eqref{E3DKdSm} and taking the limit $\epsilon\rightarrow0$ yield the near horizon metric,
\begin{equation}\label{NHE3DKdSm}
ds^{2}_{NH}=\frac{\epsilon \ell^{2}}{r_{p}^{2}-r_{-}^{2}}\left(\rho^{2}\left(\frac{r_{p}^{2}-r_{-}^{2}}{\ell^{2}r_{-}}\right)^{2} d\tau^2+d\rho^{2}\right)+r_{-}^{2}\left(d\phi-\frac{r_{p}}{\ell r_{-}}d\tau\right)^2.
\end{equation}
The following periodicity conditions, which are obtained from \eqref{NHE3DKdSm}, warrant the absence of the conical singularity on the horizon,
\begin{eqnarray}\label{pcs}
 \tau &\sim& \tau + \frac{2\pi\ell^{2}r_{-}}{r_{p}^{2}-r_{-}^{2}}=\tau+\beta, \\
  \nonumber \phi &\sim& \phi + \frac{2\pi\ell r_{p}}{r_{p}^{2}-r_{-}^{2}}=\phi+\beta\Omega=\phi+\Phi,
\end{eqnarray}
where $\Phi$ is the angular potential. From \eqref{pcs} we can read the temperature and angular velocity of the Euclidean KdS as
\begin{equation}\label{tem and velocity eucl}
  T={\left(r_{p}^{2}-r_{-}^{2}\right)\over{2\pi r_{-}\ell^{2}}},\qquad \Omega={r_{p}\over\ell r_{-}}.
  \end{equation}

\subsection{Calculation of free energy}
In this subsection we want to calculate the Euclidean on-shell action of the three dimensional de Sitter gravity given by  
\begin{equation}\label{eosa}
I_{E} = -\frac{1}{16\pi G}\,\int_{\cal M} d^3x\sqrt{g}\,(R-\Lambda) - \frac{1}{16\pi G}\,\int_{\partial \cal M} d^2x\sqrt{\lvert \gamma \rvert}\,K\,,
\end{equation}
where the second term is  Gibbons-Hawking-York (GHY) boundary term~\cite{Gibbons:1976ue,York:1972sj} and the cosmological constant reads $\Lambda=2/\ell^{2}$ where $\ell$ is the dS radius. We note that, in order to have a well-defined variational principle for 3D Einstein gravity, there are two choices for the boundary term. First one is using the usual GHY term with the usual normalization and the well-known holographic counterterm. Second one is  utilizing one half of the GHY term without  the holographic counterterm (see \cite{Detournay:2014fva} for a related discussion in Einstein gravity with negative and zero cosmological constant). The later is used in \eqref{eosa}, since it is  consistent with the results of \cite{Bagchi:2013lma} after taking the flat-space limit $\ell\rightarrow\infty$.

 The free energy is related to the canonical partition function by
\begin{equation}\label{nta}
  F = - \beta^{-1} \ln(Z)
\end{equation}
where $\beta$ is the inverse of temperature. Using Euclidean on-shell action \eqref{eosa}, the canonical partition function can be calculated by the path integral over Euclidean metric,
\begin{equation}\label{cpf}
  Z(T,\,\Omega) = \int{\cal D}g\,e^{-I_{E}[g]} = \sum_{g_0} e^{-I_{E}[g_0(T,\,\Omega)]}\,\times Z_{\rm fluct}
\end{equation}
where the Euclidean metric $g$ is written near background field $g_{0}$, $g=g_{0}+g_{fluct}$. Here, it is assumed that fluctuations $g_{fluct}$ are small enough, so that $Z_{fluct}\sim1$. The temperature $T$ and angular velocity $\Omega$ gives the correct periodicity of background field $g_{0}$.

In the background of the Euclidean metric \eqref{E3DKdSm}, the first term in \eqref{eosa} is given by $I_{E}^{EH}=-\beta (r_c^{2}-r_{-}^{2})/4G\ell^{2}$ and  the second term is $I_{E}^{GHY}=-\beta(r_{p}^{2}+r_{-}^{2}-2r_c^{2})/8 G\ell^{2}$ where $r_c$ is a cut-off on the radial coordinate $r$ and the integration of $r$ is performed in the interval $r_-\leq r\leq r_c$ i.e. outside the horizon. Hence, $I_{E}=I_{E}^{EH}+I_{E}^{GHY}$ remains  finite when $r_c\to\infty$. Finally,  the free energy of  KdS spacetime is obtained by using \eqref{nta} and \eqref{cpf} as
\begin{equation}\label{fe3DKdSs}
  F=-\frac{r_{p}^{2}-r_{-}^{2}}{8G\ell^{2}}.
\end{equation}
We note that the flat-space limit of \eqref{fe3DKdSs} is well-defined and results in the free energy of Euclidean FSC  in \cite{Bagchi:2013lma}.
\subsection{Thermodynamic quantities from calculated free energy}
It is possible to  check the validity of free energy \eqref{fe3DKdSs} by using the standard thermodynamic relations. Using \eqref{tem and velocity eucl} to  rewrite $r_{-}$ and $r_{p}$ in terms of temperature $T$ and angular velocity $\Omega$, we have
\begin{eqnarray}\label{rmrp-TOmega}
  r_{p} &=& \frac{2\pi\Omega T \ell^{3}}{\ell^{2}\Omega^{2}-1},\\
  \nonumber r_{-} &=& \frac{2\pi T\ell^{2}}{\ell^{2}\Omega^{2}-1}.
\end{eqnarray}
 Plugging them into \eqref{fe3DKdSs}, we get
\begin{equation}\label{newfe3DKdS}
  F=-\frac{\pi^{2}T^{2}}{2G\left(\Omega^{2}-\frac{1}{\ell^{2}}\right)}.
\end{equation}
Now, the standard thermodynamics gives
\begin{equation}\label{HBKdSentropy}
  S=-\frac{\partial F}{\partial T}=\frac{\pi^{2}T}{G\left(\Omega^{2}-\frac{1}{\ell^{2}}\right)}=\frac{2\pi r_{-}}{4G},
\end{equation}
which agrees with the Bekenstein-Hawking entropy. Moreover, using the first law as {$dF=-SdT-J_Ed\Omega$}, the angular momentum is  calculated as \footnote{Note that the imaginary values for the angular momentum, $J_E$, in Euclidean metric are meaningful, since angular velocity in this case is a rate of change of real angle with respect to imaginary time.  }
\begin{equation}\label{angularmKdS}
  J_E=-\frac{\partial F}{\partial \Omega}=-\frac{\pi^{2}T^{2}\Omega}{G\left(\Omega^{2}-\frac{1}{\ell^{2}}\right)^{2}}=-\frac{r_{p} r_{-}}{4G\ell}.
\end{equation}

Now, the integrated form of the first law,
\begin{equation}\label{1stlaw}
  F=U-TS-\Omega J_E,
\end{equation}
gives the internal energy,
\begin{equation}\label{internalE}
  U=-\frac{\pi^{2}T^{2}\left(\Omega^{2}+\frac{1}{\ell^{2}}\right)}{2G\left(\Omega^{2}-\frac{1}{\ell^{2}}\right)^{2}}=-\frac{r_{p}^{2}+r_{-}^{2}}{8G\ell^{2}},
\end{equation}
which is equal to $M$ by using \eqref{Euc M J} and \eqref{Euclidean radius}.
The last thermodynamics quantity but not least is  the specific heat,
\begin{equation}\label{spheatKdS}
  C=T\frac{\partial S}{\partial T}=\frac{\pi^{2}T}{G\left(\Omega^{2}-\frac{1}{\ell^{2}}\right)}=S,
\end{equation}
which is always positive.
\section{Free energy of dS and  comparison with KdS}
In this section we calculate the free energy of  dS$_3$ spacetime and compare it with the results of previous section for the free energy of KdS to infer dominant spacetime for the  three dimensional de Sitter gravity.
\subsection{Free energy of dS }
Let us start from the calculation of free energy of dS spacetime,
\begin{equation}\label{m3DdSs}
  ds^{2}=-\left(1-\frac{r^{2}}{\ell^{2}}\right) dt^{2} + \frac{1}{1-\frac{r^{2}}{\ell^{2}}} dr^{2} + r^{2} d\phi^{2}.
\end{equation}
The temperature of cosmological horizon  is obtained by using  the surface gravity $\kappa$ as $T=\kappa/2\pi=1/2\pi\ell$. The Wick rotation $t=i\tau$ results in the Euclidean metric
\begin{equation}\label{Em3DdSs}
  ds^{2}_{E}=\left(1-\frac{r^{2}}{\ell^{2}}\right) d\tau^{2} + \frac{1}{1-\frac{r^{2}}{\ell^{2}}} dr^{2} + r^{2} d\phi^{2}.
\end{equation}
The Ricci scalar and the square root of metric determinant are $R=6/\ell^2$ and $\sqrt{g}=r$, respectively. Plugging all into the first term of \eqref{eosa} and evaluating the integral gives $I_{E}^{EH}=-\beta r_c^{2}/4G\ell^{2}$ where $r_c$ is a cut-off in radial coordinate $r$. Moreover the second term of \eqref{eosa} is given by   $I_{E}^{GHY}=-\beta(\ell^{2}-2r_c^{2})/8G\ell^{2}$. Thus, $I_{E}=I_{E}^{EH}+I_{E}^{GHY}=-\beta/8G$, which is finite again. The free energy of  dS$_3$  is obtained by using \eqref{nta} and \eqref{cpf},
 \begin{equation}\label{fe3DdSs}
  F=-\frac{1}{8G}.
\end{equation}

 As usual, the entropies are given by the area law, $S=A/4G$. The free energy of dS$_3$ does not  yield the correct entropy, $F\neq-TS=-1/4G$. The inequality comes from the fact that the free energy has been calculated for whole dS spacetime with a boundary at $r\rightarrow\infty$ , rather than for a compact $S^{3}$ sphere with radius $\ell$. So, there is an internal energy $U$, which is related to the boundary term, to be taken into account. Accordingly, the suitable formula is $F=U-TS$ where $U=1/8G$. The internal energy $U$ is equal to the free energy of the region behind the horizon with the boundary at $r\rightarrow\infty$.

\subsection{   Dominant spacetime}
As seen, the free energies are in consistency with the standard thermodynamics. So, it is possible to take the final step, i.e. determining the dominant spacetime.  The two Euclidean saddle points are in the same ensemble if  they are at the same temperature.
 The temperature of dS cosmological horizon is given by $1/2\pi\ell$ and for KdS, the temperature is given by \eqref{tem and velocity eucl}. The equality of these temperatures needs that 
 \begin{equation}\label{temp condition}
 r_p^2-r_-^2=r_-\ell.
 \end{equation}
Using \eqref{temp condition}, we can write the free energy of KdS \eqref{fe3DKdSs} as
\begin{equation}\label{simp kds freee}
F=-\dfrac{r_-}{8G\ell}.
\end{equation}
Thus using \eqref{fe3DdSs} and \eqref{simp kds freee} we can conclude that:
\begin{itemize}
\item $r_->\ell$: KdS is the dominant spacetime.
\item $r_-<\ell$: dS is the dominant spacetime.  
\item $r_-=\ell$: KdS and dS coexist.
\end{itemize}

Using \eqref{Euclidean radius} and \eqref{temp condition}, we can write the above conditions in terms of Euclidean mass and angular momentum. It is not difficult to see that 
\begin{itemize}
\item   $8GM_E>3$: KdS is the dominant spacetime.
\item  $8GM_E<3$: dS is the dominant spacetime.  
\item  $8GM_E=3$: KdS and dS coexist.
\end{itemize}
However, according to \cite{Balasubramanian:2001nb}, for all KdS spacetimes we have  $8GM_E<1$ or $8GM>-1$. This means that in three dimensional de Sitter gravity, it is de Sitter spacetime which is always preferable.
This is the main result of current paper.  As mentioned in the introduction, this is not a phase transition because we fix the temperature to the temperature of dS spacetime and just determine the dominant saddle point in the partition function for a fixed temperature.

 \section{Conclusion}\label{sec:CON}
In this paper, we calculated the free energy of two spacetimes, dS and KdS, at the same temperature, and by comparing them, we found that the  dS spacetime is always dominant. 

The free energy that we obtained using on-shell action for KdS correctly results in the thermodynamic quantities of this spacetime, which is a proof of the correctness of the calculations. However, in calculating the on-shell action, we took the coefficient of the Gibbons-Hawking term half of the usual value, which made the infinite term no longer exist and counterterm action was not needed. In fact, our reason for using this unusual term is the similarity with the calculations related to  asymptotically flat spacetimes in  \cite{Bagchi:2013lma}, that the flat-space limit of all our calculations should lead to the calculations of that paper. Obviously, the complete proof of why the Gibbons-Hawking action with an unusual coefficient can establish the variational   principle in 3d dS gravity itself needs further work, which will be the subject of our future study\footnote{This study  has been done for the Einstein gravity with negative and zero cosmological constant in \cite{Detournay:2014fva}. }. The most important point in this investigation is finding  an appropriate boundary condition for the asymptotically dS spacetimes, which can open another window to de Sitter holography.

As mentioned in the Introduction, there is a local coordinate change which transforms dS and KdS spacetimes to each other. If we show coordinates of the Euclidean KdS and  Euclidean dS spacetimes by $(\tau,r,\phi)$ and $(\tau', r',\phi')$ respectively, this coordinate transformation is given by
\begin{eqnarray}
\nonumber &\tau'=\dfrac{r_p}{\ell}\tau+r_-\phi,\\
 \nonumber &r'^2=\dfrac{\ell^2}{r_p^2-r_-^2}r^2-\dfrac{\ell^2 r_-^2}{r_p^2-r_-^2},\\
 &\phi'=\dfrac{r_-}{\ell^2}\tau+\dfrac{r_P}{\ell}\phi.
\end{eqnarray}
If we assume $2\pi$ for the periodicity of $\phi$ coordinate in KdS, from above coordinate transformation we find $\Phi'$ and $\beta'$ for the periodicity of $\phi'$ and $\tau'$ where 
\begin{equation}\label{betaphip}
\beta'=2\pi r_-,\qquad \Phi'=\dfrac{2\pi r_p}{\ell}.
\end{equation}
Since $\beta$ and $\Phi$ for Euclidean KdS are given in \eqref{pcs} in terms of $r_-$ and $r_p$, we can convert their expressions  to find $r_-$ and $r_p$ and write \eqref{betaphip} as 
\begin{equation}
\beta'=4\pi^2\dfrac{\ell^2\beta}{\ell^2\Phi^2-\beta^2},\qquad \Phi'=4\pi^2\dfrac{\ell^2\Phi}{\ell^2\Phi^2-\beta^2}.
\end{equation}
As a result, we have a transformation from $(\beta,\Phi)$ to $(\beta',\Phi')$ which is similar to the modular transformation of a CFT. This can be another confirmation for the existence of a dual conformal field theory for asymptotically de Sitter spacetimes, which is still an open question in high energy physics.
\subsubsection*{Acknowledgments}
The authors would like to thank Ali Naseh for useful comments and discussions.

%\appendix

\end{document}